\documentclass[preprintnumbers]{revtex4}

\usepackage{graphicx}
\def\mpl{\ifmmode \overline M_{Pl}\else $\overline M_{Pl}$\fi}

\begin{document}
\bibliographystyle{revtex}

\preprint{}

\title{Ultrahigh Energy Neutrinos in the Light of SuperK}

\author{Sharada Iyer Dutta$^1$, Mary Hall Reno$^{2}$ and Ina Sarcevic$^1$}


\affiliation{$^1$Department of Physics, University of Arizona, Tucson, Arizona
85721\\
$^2$Department of Physics and Astronomy, University of Iowa, Iowa City,
Iowa 52242}

\date{\today}

\begin{abstract}

We propose a novel approach for studying
neutrino oscillations with extragalactic
neutrinos.  
We show that measurement of the
neutrino induced upward hadronic and electromagnetic showers
and upward muons could be used to detect
$\nu_\mu \rightarrow \nu_\tau$ oscillations.
We find significant signal to background
ratios for the hadronic/electromagnetic showers with
energies above 10 TeV to 100 TeV initiated by the
extragalactic neutrinos.
We demonstrate that a
kilometer-size neutrino telescope has a very good chance of detecting
neutrino oscillations.  
\end{abstract}

\maketitle

Recent SuperK data on atmospheric neutrinos indicate
$\nu_\mu\rightarrow \nu_\tau$ oscillations with mixing being nearly
bi-maximal \cite{superk}, in 
agreement
with previously reported results on the atmospheric anomaly
by Kamiokande \cite{kamioka} and MACRO \cite{macro} and 
consistent with limits from other experiments, e.g., CHOOZ \cite{chooz}.
Direct detection of $\nu_\tau$ appearance is extremely
difficult because  at low energies, the charged-current
cross section for producing a tau is small and the
tau has a very short lifetime.
Several long-baseline experiments with accelerator sources
of $\nu_\mu$  \cite{MINOS,K2K,ICARUS,NOE,OPERA}
have been proposed with the goal of detecting tau neutrinos
from oscillations, thus confirming the SuperK results.
Until recently, 
the only convincing evidence of neutrino oscillations involved indirect 
measurements, namely
the disappearance of the expected neutrino fluxes from the sun and 
atmospheric
neutrinos.  
A recent breakthrough in the study of neutrino oscillations came from the
SNO experiment of a direct observation of solar neutrino conversion into 
other active flavor \cite{SNO1}.  
SNO measurement of the elastic scattering 
rate is consistent with the precision elastic scattering 
measurement by SuperK \cite{fukuda}, 
and both experiments are in very good agreement with the 
theoretical prediction for the 
$^8B$ flux \cite{bachall}.

We have recently discussed complementary way of detecting neutrino 
oscillations, namely the possibility of 
a kilometer-size
neutrino telescope detecting neutrino oscillations using 
extragalactic sources of high-energy
neutrinos such as Active Galactic Nuclei (AGN) and Gamma Ray Bursts
(GRB) \cite{irs, drs00, drs01}.
The large distances involved for astrophysical sources, on the
order of one to thousands of Megaparsecs, make the next generation of
neutrino experiments potentially sensitive to neutrino mass
differences as low as $\Delta m^2\sim 10^{-17}$ eV$^2$.  
Assuming maximal mixing, over such long baselines, 
half of the neutrinos arriving at the
earth would be $\nu_\tau$'s in oscillation scenarios, the other
half being $\nu_\mu$'s. 
The effect of attenuation of the neutrino flux due to interactions
of neutrinos in the Earth is qualitatively different for 
$\nu_\mu$ and $\nu_\tau$ \cite{halzen}. Muon neutrinos are absorbed
by charged current interactions, while tau neutrinos are regenerated
by tau decays. The Earth never becomes opaque to $\nu_\tau$, though
the effect of $\nu_\tau\rightarrow \tau\rightarrow \nu_\tau$
interaction and decay processes is to degrade the energy of the incident
$\nu_\tau$. The identical spectra of $\nu_\mu$ and $\nu_\tau$ incident on the
Earth emerge after passage through the Earth with distinctly
different spectra.
The preferential penetration of $\nu_\tau$ through
the Earth is of great importance for high energy neutrino
telescopes such as AMANDA, NESTOR and ANTARES.

Here we consider $\nu_\mu$ and $\nu_\tau$ propagation
through the Earth and 
show that the energy spectrum
of the $\nu_\tau$ becomes enhanced at low energy.  
The degree of enhancement depends on
the initial neutrino flux. We consider initial fluxes $F_\nu^0\sim
E^{-n}$ for $n=1,2,3.6$, a 
GRB flux \cite{waxman} and an AGN flux \cite{stecker}.
We solve the coupled transport equations for lepton and neutrino
fluxes as indicated below.

Let $F_{\nu_{\tau}}(E,X)$ and $ F_\tau(E,X)$ be the differential energy
spectrum 
of tau neutrinos and tau
respectively at a column depth $X$ in the medium.  
Then, one can derive 
the following cascade equation for neutrinos as,
\begin{eqnarray}
\nonumber
& &\frac{\partial F_{\nu_{\tau}}(E,X)}{\partial X} =
- \frac{F_{\nu_{\tau}}(E,X)}
	{\lambda_{\nu_{\tau}}(E)}
+ \int_E^\infty dE_y
\left[\frac{F_{\nu_{\tau}}(E_y,X)}{\lambda_{\nu_{\tau}}
(E_y)}\right]
	{\frac{dn}{dE}}({\nu_{\tau}}N\rightarrow {\nu_{\tau}}X; E_y,E)
\end{eqnarray}
\begin{eqnarray}
\nonumber
&+& \int_E^\infty dE_y
\left[\frac{F_{\tau}(E_y,X)}{\rho_{\tau}^{dec}(E_y)}\right]
	{\frac{dn}{dE}}({\tau}\rightarrow {\nu_{\tau}}X; E_y,E)
+\int_E^\infty dE_y
\left[\frac{F_{\tau}(E_y,X)}{\lambda_{\tau}(E_y)}\right]
	{\frac{dn}{dE}}({\tau}N\rightarrow {\nu_{\tau}}X; E_y,E)	
\end{eqnarray}
and for taus as,
\begin{eqnarray}
\nonumber
\frac{\partial F_\tau(E,X)}{\partial X} =
- \frac{F_\tau(E,X)}{\lambda_\tau(E)}
	- \frac{F_\tau(E,X)}{\rho_\tau^{dec}(E,X,\theta)}
+ \int_E^\infty dE_y
\left[\frac{F_{\nu_{\tau}}(E_y,X)}{\lambda_{\nu_{\tau}}
	(E_y)}\right]{\frac{dn}{dE}}({\nu_{\tau}}N\rightarrow {\tau}X;
E_y,E) .
\end{eqnarray}

The first term in Eq. (1) is a loss due to the neutrino 
interactions, the second is the regeneration term due 
to the neutral current, the third term is a contribution due 
to the tau decay and the last term is the contribution due to tau
interactions. 
In Eq. (2), the first term is a loss due to tau interactions, 
the second term is a loss due to the tau decay, 
while the last term is a contribution from neutrino 
charged current interactions.  
As a practical matter, tau decays are more important than
tau interactions at the energies considered here, though interactions 
become more important at higher energies \cite{seckel}.
Here $\lambda(E)$ is the interaction length and 
$\rho_{\tau}^{dec}(E,X,\theta)$ 
is the decay length for tau. 
The charged and neutral current 
energy distributions, $dn/dE$, and the total cross section, 
$\sigma^{{tot}}_{{\nu}T}(E)$ are calculated 
taking into account recent improvements in 
our knowledge of 
the small-x behavior of the structure functions \cite{gqrs98}.

To demonstrate the importance of regeneration of tau neutrinos from
tau decays, we evaluate the tau neutrino flux for several input
neutrino spectra and compare to the attenuated $\nu_\mu$ flux.
For the incoming neutrino spectrum we use power law spectrum, i.e. 
$F_\nu^0(E) = K\left(\frac{E_0}{E}\right)^{n}$.   
In Fig. 1 we show the energy dependence of the 
the ratio of fluxes for nadir angles 
$\theta=0^0$, 
$\theta=30^\circ$ and 
$\theta=60^\circ$ for 
$\nu_\mu$ and $\nu_\tau$.  

\vspace*{-0.5cm} 
\begin{figure}[htbp]
\centerline{
\includegraphics[width=8cm,height=10cm]{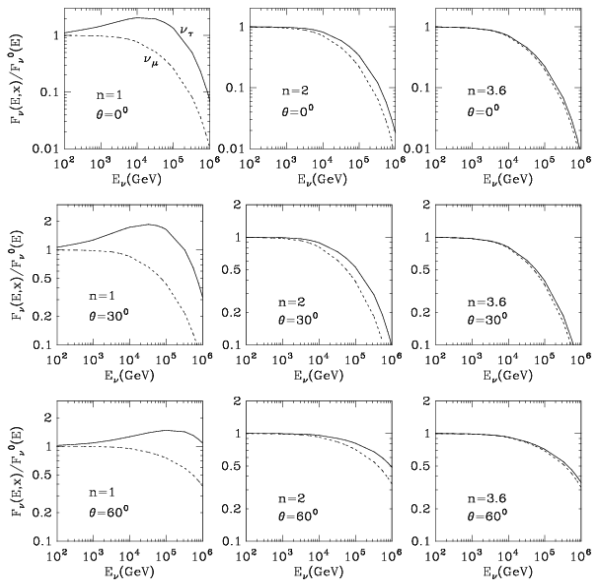}}
\vspace*{-0.9cm}
\caption{
The energy dependence of
the ratio of fluxes for nadir angles
$\theta=0^0$,
$\theta=30^0$ and
$\theta=90^0$ for
$\nu_\mu$ and $\nu_\tau$ assuming
$F_\nu^0(E) \sim E^{-n}$ with $n=1,2$ and $3.6$.} 
\label{fig1}
\end{figure}
\vspace*{-0.1cm}
For small nadir angles, $\theta=0$ and $30^\circ$
and 
$F_\nu^0(E)\sim 1/E$ we find that enhancement of tau neutrinos is in the 
energy range of $10^2$ GeV and $10^5$ GeV, while for 
$\theta=60^\circ$, 
the enhancement extends up to $10^6$ GeV.  In contrast 
the $\nu_\mu$ flux is attenuated for all the nadir angles.  When the 
incoming flux is steeper, $n=2$, the $\nu_\tau$ flux appears to be attenuated 
at high energies, although less than the $\nu_\mu$ flux.    For $n=3.6$, 
the energy dependence of these two fluxes is very similar, they are both 
reduced at high energies, and the effect is stronger for smaller nadir 
angle, since in this case the column depth is larger and there are more 
charged current interactions possible.  
In case of the AGN quasar model \cite{stecker}, for example, 
we find that the $\nu_\tau$ flux is a factor of $2$ to $2.5$ times larger than 
the input flux, for nadir angle, $\theta=0$ and $E=10^2-10^4$ GeV\cite{irs}.  
For larger angles, the effect is smaller. 

The appearance of high energy tau neutrinos due to
$\nu_\mu \rightarrow \nu_\tau$ oscillations of extragalactic neutrinos
can be observed by measuring the neutrino induced 
upward hadronic and electromagnetic showers 
and upward muons.  
Charged-current interactions of the upward tau neutrinos
below and in the detector, and the subsequent tau decay
create
muons or hadronic and electromagnetic showers.  The background
for these events are muon neutrino and electron neutrino charged-current and
neutral-current interactions, where in addition to extragalactic neutrinos, we
also include the background from atmospheric neutrinos.

In Fig. 2a) we show that 
tau neutrinos give significant contributions
to upward hadronic/EM showers, 
signaling the
$\nu_\tau$ appearance.
The solid lines show the event rates including contributions for $\nu_\tau,\ 
\nu_\mu$ and $\nu_e$. The dashed lines are from $\nu_\mu+\nu_e$, assuming
equal neutrino and antineutrino fluxes of each flavor. 
We note 
that in the case of the $E^{-1}$ flux,
the contributions from tau neutrinos are large, a factor of
4 times larger than the muon neutrino plus electron neutrino
contribution at zero nadir angle.
For horizontal showers, the enhancement factor is smaller, about 2 for
all the energy thresholds that we consider.
Similarly, for $E^{-2}$ flux, the tau neutrino contribution is
a factor of 1.7 times larger than the muon neutrino plus electron neutrino
contributions for upward
showers.  
The rates for AGN quasar model \cite{stecker}, 
at zero nadir angle, 
are comprised of 60\%  tau neutrino induced events, decreasing to about 40\%
tau neutrino induced events for
horizontal showers, 
translating to  25-80 shower events for $E_{\rm shr}^{\rm min}=10$ TeV and 6-%
45 events for
$E_{\rm shr}^{\rm min}=100$ 
TeV with negligible atmospheric background \cite{drs00}.  

We note that for the  $E^{-1}$ flux,
the shower event rates for $\nu_\tau+\nu_\mu+\nu_e$  are
a factor of 3.3-3.7 larger than in the $\nu_\mu+\nu_e$ contributions
in the no-oscillation scenario
for $E_{\rm shr}^{\rm min}=1-100$ TeV for $\theta=0^{\circ}$.
They are a factor of 1.6 enhanced for the horizontal shower rate.
For the $E^{-2}$
flux, the enhancement is a factor
of 1.4-1.6 relative to the 
$\nu_\mu+\nu_e$ no-oscillation
shower rate for $E_{\rm shr}^{\rm min}=1-100$ TeV.  
In the case of AGN models, if one assumes oscillations, the shower event
rates are factor of 1.8-2.1 larger at zero nadir angle, decreasing
to 1.5 for nearly horizontal showers \cite{drs00}.  
Given the uncertainties in the normalizations
of the extragalactic neutrino fluxes, combining muon rates and hadronic/EM
rates offer the best chance to test the $\nu_\mu\rightarrow \nu_\tau$
oscillation hypothesis.

As concluded in earlier work \cite{gqrs96,gqrs98}, in general,
an energy threshold of between 10 TeV and  100 TeV for
upward muons and showers
is needed in order to reduce the background from atmospheric neutrinos.
We find that diffuse AGN neutrino fluxes, 
as well as
neutrinos from GRBs can be used to detect tau appearance.
By measuring upward showers with
energy threshold of 10 TeV, and upward muons, the event rates exceed
the atmospheric
background and are about a factor of 1.5-2 larger than in the no-oscillation
scenario \cite{drs00}.

To determine the relative
enhancement of the hadronic/EM signal compared to the muon signal,
one can compare the rates for horizontal events, where tau neutrino pileup
is small.
For example,
Fig. 2b) shows a clear distinction between oscillation and
no-oscillation scenarios, even in directions near horizontal, where there is
no pileup.  Even with a small tau neutrino pileup, the
oscillation scenario can be distinguished from the
no-oscillation scenario.

\begin{figure}[htbp]
\centerline{
\includegraphics[width=5cm]{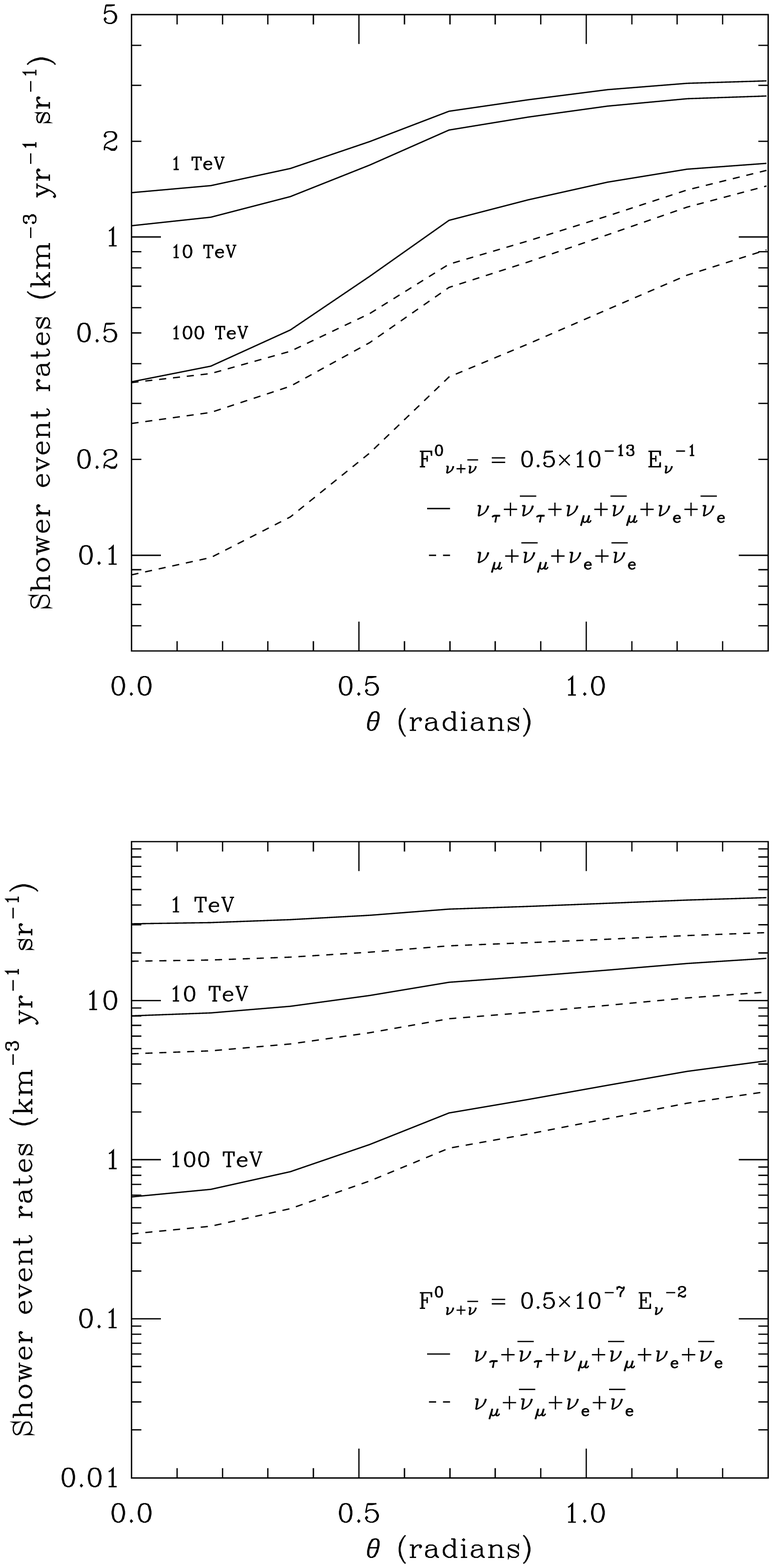}
\hspace*{15mm}
\includegraphics[width=8cm]{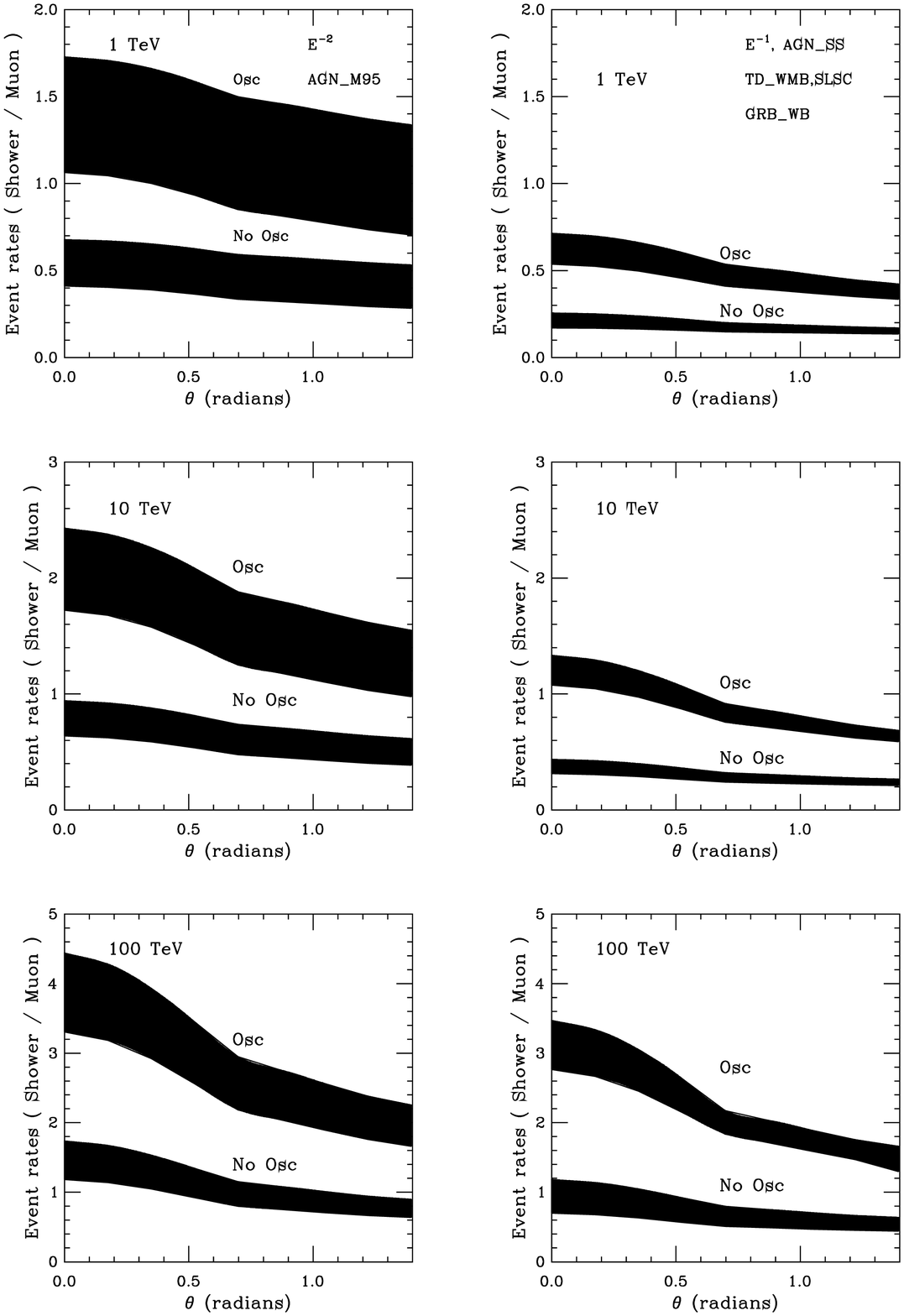}}
\vspace*{0.1cm}
\caption{
a) 
Hadronic/EM event rates as a
function of nadir angle for $E_{\rm shr}^{\rm min}=$1 TeV,10 TeV and 100 TeV.
Hadronic/EM event rates from $\nu_\tau$ (solid line) compared
hadronic/EM event rates from $\nu_\mu$ plus $\nu_e$ (dashed line) for
$E^{-1}$ and $E^{-2}$.  b)
Ratio of Hadronic/EM event rate to muon event
rate for the oscillation (upper shaded area) and no-oscillation
(lower shaded area) scenarios
as a function of nadir
angle for threshold energies of 1 TeV, 10 TeV and
 100 TeV for the indicated fluxes.
}

\label{fig2}
\end{figure}

The detection of
$\nu_\mu \rightarrow \nu_\tau$ oscillations with a point source might also
be possible.  With the
resolution for the planned neutrino telescopes of $2^{\circ}$, the atmospheric
background is reduced by $3.8 \times 10^{-3}$.
For upward showers, this gives less than 1 event per year for
$E_{\rm shr}^{\rm min}=1$ TeV, and even less
for higher energy thresholds.  Thus,
if the point source has a flat spectrum,
$F_{\nu+\bar{\nu}} = 10^{-16} E^{-1}$,
then one would be able to detect tau neutrinos by measuring upward
showers with $E_{\rm shr}^{\rm min}=1$ TeV.
In the more realistic case, when the point source has a steeper spectrum
($E^{-2}$),
such as Sgr A* \cite{markoff98}, a normalization of 
$10^{-7}$/(cm$^2$s\,sr\,GeV)
would be sufficient for the
detection of tau neutrinos with threshold of 1 TeV. Time correlations with
variable point sources would further enhance
the signal relative to the background.

We have demonstrated that extragalactic sources of neutrinos can be used
as a very-long baseline experiment, providing a source of tau neutrinos
and opening up a new frontier in studying neutrinos oscillations.

\vskip 0.1true in

\leftline{\bf Acknowledgements}

This work has been supported in part
by the DOE under Contract
DE-FG02-95ER40906 and 
in part by
National Science Foundation Grant No.
PHY-9802403.

\end{document}